\documentclass[a4paper,11pt]{article}
\usepackage{graphicx,amssymb,amsmath,amsfonts,epsfig}
\usepackage[usenames,dvipsnames]{color}
\usepackage{graphicx}  
\usepackage{amsmath}   
\usepackage[compress]{cite}
\usepackage{amssymb}   
\usepackage{bm} 
\usepackage{dcolumn}
\usepackage{color}
\usepackage[dvipsnames]{xcolor}
\usepackage{mathrsfs}
\usepackage{gensymb}
\usepackage{float}
\usepackage{amsfonts}
\usepackage{varioref}
\usepackage{textcomp}
\usepackage{enumerate}
\usepackage{tikz}
\usepackage{multicol}
\usepackage{soul}
\usepackage{notoccite}
\usepackage{graphicx}
\usepackage{subfigure}
\title{Zero Mass limit of Kerr-MOG Black Hole Equals Wormhole}
\author{Parthapratim Pradhan\footnote{\color{Blue}pppradhan77@gmail.com}~$^{1}$ and 
John W. Moffat\footnote{\color{Blue}jmoffat@perimeterinstitute.ca}~$^{2,3}$\\
{$^{1}$\small{\it Department of Physics, Hiralal Mazumdar Memorial College For Women, Dakshineswar, Kolkata-700035, India}}\\
{$^{2}$\small {\it Perimeter Institute for Theoretical Physics, Waterloo, Ontario N2L 2Y5, Canada}}\\
{$^{3}$\small{\it Department of Physics and Astronomy, University of Waterloo, Waterloo, Ontario N2L 3G1, Canada}}}

\date{}

\begin{document}

\maketitle

\begin{abstract}
It has been argued in existing literature that the zero mass limit of Kerr spacetime corresponds to either flat Minkowski 
spacetime or a wormhole exhibiting a locally flat geometry. In this study, we examine that the zero mass limit of the 
Kerr-MOG black hole is equivalent to a wormhole. Moreover, we derive the Kerr-Schild form of the Kerr-MOG black hole 
through specific coordinate transformations. We further investigate the physical and topological 
characteristics of the Kerr-MOG black hole within the framework of modified gravity. Our analysis also includes a 
discussion of the wormhole using cylindrical coordinates, which comprises two distinct coordinate patches. 
Furthermore, we extend our analysis to the Kerr-Newman black hole and show that the \emph{zero mass limit of the 
Kerr-Newman black hole does not yield a wormhole}. However, if we impose an additional criterion such that
\emph{both the mass parameter and the charge parameters are  equal to zero}, then the Kerr-Newman black hole will 
be a wormhole.   
\end{abstract}

\section{Introduction}
In a recent work\cite{GW17}, it has been shown that the zero mass limit of the Kerr spacetime is not flat Minkowski 
spacetime. Instead, it locally becomes a flat static wormhole spacetime containing a conical singularity of the Ricci 
tensor along a ring. The singularity generated in the spacetime should be interpreted as an effect of a singular matter 
source- a negative tension cosmic string loop\cite{GW17,Vol17}. Also, they showed that the zero mass limit of the Kerr-AdS 
or Kerr-de-Sitter spacetime is a locally AdS wormhole supported by a negative tension ring. 

Motivated by this work, we show that the zero mass limit of the Kerr-MOG black hole is equal to a wormhole. To do that 
we first derive the \emph{Kerr-Schild form of the Kerr-MOG black hole} by using some coordinate transformations. 
Moreover, we examine the physical and topological feature of the Kerr-MOG black hole in modified gravity~(MOG). We also 
discuss the wormhole by using cylindrical coordinates, which consists of two coordinate patches to cover the
whole geometry. Furthermore, we examine the geodesics along the symmetry axis\cite{c1}. Also, we discuss the maximal analytic 
extension of it. Finally, we extend our study to the \emph{ Kerr-Newman black hole, showing that the zero mass limit 
of the Kerr-Newman black hole is not a wormhole}. It is possible only when both \emph{mass and charge parameters 
approach to zero value}.

There has been a resurgence of interest in the physics of MOG~\cite{mf,mf1,mf2,mf3,mf4,mf5,mf6,mf7,mf8,mf9,mf10}. The 
primary focus of this interest lies in the potential of MOG theory to elucidate the large-scale behavior of gravitational
fields, suggesting it may serve as an alternative to Einstein's general relativity by eliminating the need to postulate 
directly unobserved dark matter.

A key characteristic of MOG theory is the presence of a massive vector field, which interacts with the spacetime metric 
to influence the gravitational field. This interaction implies that MOG black holes possess a gravitational charge that 
is directly proportional to their mass. Additionally, the gravitational constant $G$ is represented 
as a scalar field that varies with the MOG parameter $\alpha$, which quantifies the divergence of MOG from GR. The MOG
parameter is mathematically defined as $\alpha = \frac{G - G_{N}}{G_{N}}$, where $G_{N}$ denotes the 
Newtonian gravitational constant. A fundamental tenet of MOG theory posits that the gravitational charge is proportional 
to the square root of the MOG parameter, expressed as $Q_g = \sqrt{\alpha G_{N}}M$.

It is generally accepted that a black hole has negligible electric charge~\cite{Wald}.
The electromagnetic repulsion in compressing an electrically charged mass is significantly greater than the gravitational 
attraction by about 40 orders of magnitude. Therefore, it is not expected that, as with all astrophysical bodies including 
neutron stars, black holes with a significant electric charge will form in nature. The gravitational charge $Q_g$ is 
proportional to positive mass $M$ and accumulates with increasing mass, while the positive and negative charges of 
protons and electrons, respectively, rapidly neutralize the net electrical charge in astrophysical bodies and black holes.

The structure of the paper is outlined as follows. In Sec. 2, we review the Kerr solutions of the vacuum Einstein 
equations. In Sec. 3, we present our primary finding, which is the Kerr-Schild representation of the Kerr-MOG black 
hole, achieved through specific coordinate transformations. In  Sec. 4, we discuss the wormhole geometry in cylindrical 
coordinates. In Sec. 5, we derive the zero mass limit of the Kerr-MOG black hole demonstrating that it is a wormhole. 
In Sec. 6, we extend our analysis to the Kerr-Newman black hole to show that the zero mass limit of it is not a 
wormhole. However, it becomes a wormhole when both the mass parameter and the electric charge parameter tend to 
zero. Sec. 7 contains our concluding remarks.

\section{Review of the Kerr-Schild Form of the Kerr Black Hole}


Before advancing to the derivation of the Kerr-Schild~\cite{ks65} form of the Kerr-MOG black hole in the following 
section, it is important to briefly review the Kerr-Schild form of the Kerr black hole. The metric ansatz utilized 
by Kerr and Schild is distinguished by the following structure
\begin{eqnarray}
ds^2 &=& g_{ab} dx^{a}dx^{b} \equiv \eta_{ab}dx^{a}dx^{b}- 2 h k_{a}k_{b} dx^{a}dx^{b} ~.\label{ksm1}
\end{eqnarray}
where $\eta_{ab}$ is the Minkowski metric, $h$ is a scalar function and $k_{a}$ is a null
vector with respect to both metrics $g_{ab}$ and $\eta_{ab}$ so that we have
\begin{eqnarray}
g_{ab} k^{a} k^{b} &=& \eta_{ab} k^{a} k^{b} = 0,\,\,\ g^{ab} = \eta^{ab} + 2h k^a k^b ~.\label{ksm2}
\end{eqnarray}
It should be noted that, because of $g_{ab}g^{bc}=\delta^{c}_{a}$, the determinant of the metric is independent 
of $h$. This implies that $|g_{ab}|=|\eta_{ab}|$, where $\eta_{ab}=[-1,1,1,1]$.
The ansatz Eq.(\ref{ksm1}) and Eq.(\ref{ksm2}) were first introduced by Trautman~\cite{trau62}. Sometimes the ansatz 
is called Kerr-Schild-Trautman ansatz~\cite{stephani}. If  $(x^{0}=t, x^{1}=x, x^{2}=y, x^{3}=z)$ 
are the Cartesian coordinates for Minkowski spacetime and introducing the complex null coordinates~
$(\zeta,\bar{\zeta},u,v)$~\cite{dks69,kerrwilson78}[{See also \cite{gurses75,xanthopulos78}}] by
\begin{eqnarray}
\zeta &=& \frac{x+iy}{\sqrt{2}},\,\, \bar{\zeta}=\frac{x-iy}{\sqrt{2}} \\
u &=& \frac{t-z}{\sqrt{2}},\,\, v = \frac{t+z}{\sqrt{2}} ~.\label{ksm3}
\end{eqnarray}
then the metric~(\ref{ksm1}) becomes
\begin{eqnarray}
ds^2 &=& 2(d\zeta d\bar{\zeta}-du dv)- 2 h k_{a}k_{b}dx^{a}dx^{b} ~.\label{ksm4}
\end{eqnarray}
where {$\eta_{ab}dx^{a}dx^{b}=2(d\zeta d\bar{\zeta}-du dv)$}.  This metric can be rewritten as 
\begin{eqnarray}
ds^2 &=& 2d\zeta d\bar{\zeta}-2du dv- 2 h (k_{a}dx^{a})^2 ~.\label{ksm5}
\end{eqnarray}
A general field of real null vectors in flat Minkowski spacetime~\cite{cox76} 
is defined as 
\begin{eqnarray}
k_{a}dx^{a} &=&-[du+Y\bar{Y}dv+\bar{Y}d\zeta+Yd\bar{\zeta}]
\end{eqnarray}
where $Y$ is an arbitrary complex function of coordinates.  One can introduce the tetrad 
frame $\omega^{1}=d\zeta+Y dv$, $\omega^{2}=\bar{\sigma^{1}}=d\bar{\zeta}+\bar{Y} dv$, 
$\omega^{3}\equiv k_{a}dx^{a}=-[du+Y\bar{Y}dv+\bar{Y}d\zeta+Yd\bar{\zeta}]$ and  
$\omega^{4}=dv+h\sigma^{3}$.  
Then the metric~(\ref{ksm5}) becomes
\begin{eqnarray}
ds^2 &=& 2 \left(\omega^{1} \omega^{2}-\omega^{3}\omega^{4}\right). 
\end{eqnarray}
The dual frame is defined by $\mathbf{e}_{1}=\partial_{\zeta}-\bar{Y}\partial_{u}$, 
$\mathbf{e}_{2}=\bar{\mathbf{e}_{1}}=\partial_{\bar{\zeta}}-Y\partial_{u}$, 
$\mathbf{e}_{3}=\partial_{u}-h\mathbf{k}$ and 
$\mathbf{e}_{4}=\mathbf{k}=k^{a}\partial_{a}
=Y\bar{Y}\partial_{u}+\partial_{v}-Y\partial_{\zeta}-\bar{Y}\partial_{\bar{\zeta}}$.    
The tetrad formalism is not presented in 
detail, as it can be found in Refs. \cite{ks65,dks69,bgk10}. Utilizing these properties, Kerr successfully derived 
the Kerr solutions for the vacuum Einstein equations in 1963~\cite{kerr63}.
\section{The Kerr-Schild Form of the Kerr-MOG Black Hole}
We will begin our main analysis by considering the metric of the Kerr-MOG black hole~\cite{mf,mf1,mf2,mf3,mf4} as 
\begin{eqnarray}
ds^2 &=& - \frac{\Upsilon_{r}}{\varrho^2} \, \left[dt-a\sin^2\theta d\phi \right]^2+\frac{\sin^2\theta}{\varrho^2}
\,\left[(r^2+a^2) \,d\phi-a dt\right]^2
+\varrho^2 \, \left[\frac{dr^2}{\Upsilon_{r}}+d\theta^2\right] ~.\label{ks1}
\end{eqnarray}
where
\begin{eqnarray}
\varrho^2 & \equiv & r^2+a^2\cos^2~\theta, \,\,
\Upsilon_{r}  \equiv  r^2-2G_{N}(1+\alpha)Mr+a^2 + G_{N}^2 \alpha(1+\alpha) M^2  ~.\label{ks2},
\end{eqnarray}
where $M$ is the mass parameter {defined in} Ref.~\cite{mf8}. The above metric is written 
in Boyer-Lindquist coordinates~$(t,r,\theta,\phi)$. The metric has some basic features. The horizon function 
$\Upsilon_{r}$ depends on the spin parameter, mass parameter and MOG parameter. It is a stationary and 
axisymmetric solution of Einstein's equations.

It should be noted that the ADM mass parameter and the angular momentum parameter 
{should be defined as}
${\cal M}=(1+\alpha)M$ and $J=a G_{N}{\cal M}$~\cite{ps}. 
Thus, the horizon function becomes 
\begin{eqnarray}
\Upsilon_{r} &=& r^2-2G_{N}{\cal M}r+a^2 + \frac{\alpha}{(1+\alpha)} G_{N}^2{\cal M}^2 
\end{eqnarray}
The above mentioned black hole consists of two horizons, namely, the event horizon~($r_{H}$) and the Cauchy 
horizon~($r_{C}$). They are described as 
\begin{eqnarray}
r_{H} &=&  G_{N}{\cal M} + \sqrt{\frac{G_{N}^2{\cal M}^2}{1+\alpha}-a^2},\,\,\,
r_{C} =  G_{N}{\cal M} - \sqrt{\frac{G_{N}^2{\cal M}^2}{1+\alpha}-a^2} ~.\label{ks4}
\end{eqnarray}
It should be noted that when $\alpha=0$, one obtains the horizon radii of Kerr black hole. The black hole 
solution does exist when $\frac{G_{N}^2{\cal M}^2}{1+\alpha} > a^2$. When $\frac{G_{N}^2{\cal M}^2}{1+\alpha}=a^2$, one 
gets the extremal black hole. When $\frac{G_{N}^2{\cal M}^2}{1+\alpha}<a^2$, one finds the naked singularity situation.

For our purpose, the above metric could be rewritten as 
\begin{eqnarray}
ds^2 &=& -dt^2+\varrho^2 \, \left(\frac{dr^2}{\Upsilon_{r}}+d\theta^2\right)+(r^2+a^2)\sin^2\theta d\phi^2
+\frac{\Pi_{\alpha}}{\varrho^2}\,\left(dt-a\sin^2\theta d\phi \right)^2 ~.\label{ks5}
\end{eqnarray}
where 
\begin{eqnarray}
\Pi_{\alpha} &=& \left(2G_{N}{\cal M}r-\frac{\alpha}{1+\alpha} G_{N}^2{\cal M}^2\right).
\end{eqnarray}
Interestingly, when ${\cal M}=0$, {one obtains} the metric in the following form
\begin{eqnarray}
ds^2 &=& -dt^2+\left(r^2+a^2\cos^2\theta\right)\frac{dr^2}{r^2+a^2}+\left(r^2+a^2\cos^2\theta\right) d\theta^2+
(r^2+a^2)\sin^2\theta d\phi^2 ~.\label{ks6}
\end{eqnarray}
Indeed, it is the metric of Minkowski spacetime in spheroidal coordinates or it is a flat space in oblate 
spheroidal cordinates. This derivation stems from the Minkowski spacetime. We start with the Minkowski spacetime 
{which could be} represented as follows:
\begin{eqnarray}
ds^2 &=& -dt^2+dx^2+dy^2+dz^2 ~.\label{ks7}
\end{eqnarray}
This equation describes the metric of Minkowski space in Cartesian coordinates~$(t,x,y,z)$.
To convert this metric into cylindrical coordinates, we apply the transformation
\begin{eqnarray}
x=\rho \cos\phi,\,\, y=\rho \sin\phi, \label{ks7.2} 
\end{eqnarray}
 resulting in the spacetime metric:
\begin{eqnarray}
ds^2 &=& -dt^2+d\rho^2+\rho^2d\phi+dz^2 ~.\label{ks7.1}
\end{eqnarray}
Next, to derive the metric in oblate spheroidal coordinates, we utilize the coordinate transformation 
$\rho=\sqrt{r^2+a^2} \sin\theta$ and $z=r\cos\theta$ (where $r\in [0,+\infty)$ and $\theta \in [0,\pi]$). 
This leads us to the metric as expressed in Eq.~(\ref{ks6}). Now we will calculate the Jacobian of the 
coordinate transformation
\begin{eqnarray}
J=\Big|\frac{\partial (x,y,z)}{\partial(r,\theta,\phi)}\Big|=\left(r^2+a^2\cos^2\theta
\right)\sin\theta.
\end{eqnarray}
It disappears at the point where $r=0$ and $\theta=\frac{\pi}{2}$, therefore for $\rho^2\equiv x^2+y^2=a^2$, 
and $z=0$. This situation represents a ring with a radius of $a$ situated in the equatorial plane. Consequently, the 
coordinates $(r,\theta,\phi)$ encompass the entirety of Minkowski spacetime, with the exception 
of the $z$ axis and the aforementioned ring. 

Let us now examine the concept of coordinated singularity at the ring. We have the equation
\begin{eqnarray}
{\frac{x^2+y^2}{r^2+a^2 }+\frac{z^2}{r^2}} &=& \frac{\rho^2}{r^2+a^2 }+\frac{z^2}{r^2} = 1. 
\end{eqnarray}
This indicates that lines of constant $r$ are oblate (half-) ellipses in the $(\rho,z)$ plane, and while hyperbolas 
of constant $\theta$ are orthogonal to these ellipses~[See Fig.~1 of Ref. (\cite{GW17})]. Consequently, two coordinate 
charts,  $(\rho_{+},z_{+})$ and  $(\rho_{-},z_{-})$
are required to adequately represent the geometry described by equation (\ref{ks6}) with $r\in (-\infty,+\infty)$. 
In the $r\rightarrow 0$ limit, the ellipses collapse to the segment of the
$\rho$ axis, denoted as 
\begin{eqnarray}
{\cal I}=(\rho \in [0,a], z=0), \label{ks6.1}.
\end{eqnarray}
Mean while $\theta$ exhibits a discontinuity 
across this segment, as shown by 
\[\lim_{z\rightarrow \pm 0} \cos\theta=\pm \sqrt{1-\frac{\rho^2}{a^2}}\,\, \mbox{if}\,\,\rho\leq a,\] and 
\[\lim_{z\rightarrow  0} \cos\theta=0\,\,\mbox{if}\,\, \rho \geq a.\]
This can be interpreted as follows. The inverse coordinate transformation $(\rho, z) \rightarrow (r,\theta)$ is 
as follows:
\begin{eqnarray}
\Big(r+ia\cos\theta\Big)^2 &=& r^2-a^2\cos^2\theta+2iar\cos\theta\\
&=&\rho^2+z^2-a^2+2iar\cos\theta\\
&=&\rho^2+z^2-a^2+2iaz\\
&=&\rho^2+(z+ia)^2, 
\end{eqnarray}
Hence, $r+ia\cos\theta$ has a branch point at $(\rho, z)= (a,0)$ and the segment in Eq.~(\ref{ks6.1}) corresponds 
to the branch cut position. Selecting a single branch of the square root results in a non-negative real part, denoted as 
$r$, while the imaginary component, represented by $a\cos\theta$, becomes inherently discontinuous across the cut. 
Consequently, the coordinates $(r, \theta)$ exhibit a discontinuity within the disk of radius $a$ 
situated in the equatorial plane.

{Now it is clearly} evident from Eq.~(\ref{ks5}) that when $\Upsilon_{r}=0$, the metric becomes 
singular. To determine if this is merely a coordinate singularity or indicative of a different type of 
singularity, it is necessary to perform a following coordinate transformation:
\begin{eqnarray}
dv &=& dt +\frac{r^2+a^2}{\Upsilon_{r}} dr \\
d\tilde{\phi} &=& d\phi+\frac{a}{\Upsilon_{r}} dr 
\end{eqnarray}
After this transformation, the new metric becomes
$$
ds^2= -\left(1-\frac{\Pi_{\alpha}}{\varrho^2}\right)dv^2 +2dv dr+ \varrho^2 d\theta^2-2a \sin^2\theta dr d\tilde{\phi}
$$
\begin{eqnarray}  
+\frac{\sin^2\theta}{\varrho^2}\,\left[(r^2+a^2)^2-a^2 \sin^2\theta \Upsilon_{r}\right] d\tilde{\phi}^2 
-2a\sin^2\theta\left(\frac{\Pi_{\alpha}}{\varrho^2}\right) dv d\tilde{\phi} ~.\label{ks8}
\end{eqnarray}

It is {evident} from this equation that the metric remains finite at $\Upsilon_{r}=0$, indicating 
that it represents 
merely a coordinate singularity. The crucial question is how to eliminate this singularity. The 
coordinates $(v,r,\theta,\tilde{\phi})$ used to express the metric are referred to as Kerr coordinates. 
The benefits of these coordinates over Boyer-Lindquist coordinates include: (i) the metric does not 
exhibit singular behavior at $\Upsilon_{r}=0$, (ii) the $g_{rr}$ component equals zero, and (iii) the 
components $g_{vr}$ and $g_{\tilde{\phi}r}$ are independent of $r$. This metric can be reformulated 
in a simpler manner.
$$
ds^2=-dv^2 +2dv dr+ \varrho^2 d\theta^2-2a \sin^2\theta dr\, d\tilde{\phi}+
(r^2+a^2)\sin^2\theta\, d\tilde{\phi}^2
$$
\begin{eqnarray} 
+\frac{\Pi_{\alpha}}{\varrho^2} \left(dv-a\sin^2\theta d\tilde{\phi}\right)^2 ~.\label{ks9}
\end{eqnarray}
If we {could} write down the metric explicitly for the time coordinate then one {should}
define  $\tilde{t}\equiv v-r$ and one obtains 
$$
ds^2=-d\tilde{t}^2 +dr^2+ \varrho^2\, d\theta^2-2a \sin^2\theta dr\, d\tilde{\phi}+
(r^2+a^2)\sin^2\theta\, d\tilde{\phi}^2
$$
\begin{eqnarray} 
+\frac{\Pi_{\alpha}}{\varrho^2} \left(d\tilde{t}+dr-a\sin^2\theta d\tilde{\phi}\right)^2 
~.\label{ks10}
\end{eqnarray}
{The above equation indicates} that there is another coordinate singularity that occurs at 
$\varrho^2=0$, specifically 
at $r=0$ and $\theta=\frac{\pi}{2}$. We will soon demonstrate that this coordinate singularity is, in fact, a curvature 
singularity. To {understand} the curvature singularity, it is essential to select a coordinate system that 
remains well-defined at $r=0$ and does not result in a divergence of the metric. Kerr and Schild have introduced such 
coordinates that are valid at $r=0$. For our analysis, we {should} define the Kerr-Schild coordinates 
as $(\tilde{t}, x, y, z)$.
\begin{eqnarray}
x &=& (r^2+a^2)^\frac{1}{2} \sin\theta\cos\left(\tilde{\phi}+\tan^{-1}\frac{a}{r}\right)\label{ks11a} \\
y &=& (r^2+a^2)^\frac{1}{2} \sin\theta\sin\left(\tilde{\phi}+\tan^{-1}\frac{a}{r}\right)\label{ks11b}\\
z &=& r\cos\theta ~.\label{ks11c}
\end{eqnarray}
In this coordinate frame, we can easily derived some useful relations 
\begin{eqnarray}
x^2+y^2 &=& \left(r^2+a^2 \right) \sin^2\theta \label{ks1.5} \\
z^2 &=& r^2 \cos^2\theta
\end{eqnarray}
and 
\begin{eqnarray}
\frac{x^2+y^2}{r^2+a^2 }+\frac{z^2}{r^2} &=& 1, ~\label{ks1.1} \\
\frac{x^2+y^2}{a^2 \sin^2\theta }-\frac{z^2}{a^2 \cos^2\theta} &=& 1 ~.\label{ks1.2}
\end{eqnarray}
To derive the metric in Kerr-Schild coordinates and in order to understand the singularity 
structure of Kerr-MOG black hole, we {could define}
\begin{eqnarray}
 \varsigma &=& \tan^{-1}\frac{a}{r}.
\end{eqnarray}
Then one finds 
\begin{eqnarray}
\cos^2\varsigma &=& \frac{r^2}{r^2+a^2}\\
\sin^2\varsigma &=& \frac{a^2}{r^2+a^2}
\end{eqnarray}
Using these relations,  the Eq.(\ref{ks11a}), Eq.(\ref{ks11b}) and Eq.(\ref{ks11c}) {should} 
be rewritten as
\begin{eqnarray}
x &=&  \sin\theta\left(r\cos\tilde{\phi}-a \sin\tilde{\phi}\right)\\
y &=&  \sin\theta\left(r\sin\tilde{\phi}+a \cos\tilde{\phi}\right) \\
z &=& r \cos\theta ~.\label{ks12}
\end{eqnarray}
Differentiating these coordinates and adding the square of the differentials, we obtain: 
\begin{eqnarray}
dx^2+dy^2+dz^2 &=& dr^2+ \varrho^2\, d\theta^2-2a \sin^2\theta dr\, d\tilde{\phi}
+(r^2+a^2)\sin^2\theta\, d\tilde{\phi}^2 ~.\label{ks13}
\end{eqnarray}
Using the differentials $dx$, $dy$ and $dz$ and after a long algebraic calculation we find
\begin{eqnarray}
\frac{r(xdx+ydy)}{r^2+a^2}-\frac{a(xdy-ydx)}{r^2+a^2}+\frac{zdz}{r} 
&=& dr-a\sin^2\theta d\tilde{\phi}~.\label{ks14}
\end{eqnarray}
Therefore,
\begin{eqnarray} 
\left(d\tilde{t}+dr-a\sin^2\theta d\tilde{\phi}\right)^2 &=& 
\left[d\tilde{t}+\frac{r(xdx+ydy)}{r^2+a^2}-\frac{a(xdy-ydx)}{r^2+a^2}+\frac{zdz}{r}\right]^2 
~.\label{ks15}
\end{eqnarray}
In Kerr-Schild coordinates 
\begin{eqnarray}
\varrho^2 &=& \frac{r^4+a^2z^2}{r^2}~.\label{ks16}
\end{eqnarray}
thus 
\begin{eqnarray}
\frac{\Pi_{\alpha}}{\varrho^2} &=& \frac{r^2\Pi_{\alpha}}{r^4+a^2z^2}~.\label{ks17}
\end{eqnarray}
{Using } the Eq.~(\ref{ks13}), Eq.~(\ref{ks14}),  Eq.~(\ref{ks15}) and  Eq.~(\ref{ks17}), the 
Eq.~(\ref{ks10}) {could} be rewritten as
\begin{eqnarray} 
ds^2=-d\tilde{t}^2 +dx^2+dy^2+dz^2+\frac{r^2\Pi_{\alpha}}{r^4+a^2z^2}
\left[d\tilde{t}+\frac{r(xdx+ydy)}{r^2+a^2}-\frac{a(xdy-ydx)}{r^2+a^2}+\frac{zdz}{r}\right]^2 
~.\label{ks18}
\end{eqnarray}
This is the Kerr-Schild form of Kerr-MOG black hole. This is {one of the} \emph{key} result 
of our work.  {The above equation could be written in a compact form as}
\begin{eqnarray} 
ds^2=ds_{\ast}^2+h (k_{\nu}dx^{\nu})^2
\end{eqnarray}
where  $ds_{\ast}^2=-d\tilde{t}^2 +dx^2+dy^2+dz^2$ is the flat Minkowski metric and 
the scalar field $h$ is given by
\begin{eqnarray} 
h &=& \frac{r^2\Pi_{\alpha}}{r^4+a^2z^2}~.\label{ks19}
\end{eqnarray}
Moreover, $k$ is a  null vector of the following form:
\begin{eqnarray} 
k_{\nu}dx^{\nu} &=& -\left(d\tilde{t}+\frac{r(xdx+ydy)}{r^2+a^2}-\frac{a(xdy-ydx)}{r^2+a^2}+\frac{zdz}{r}\right)
~.\label{ks20}
\end{eqnarray}
In Eq.~(\ref{ks18}) the radial coordinate $r$ is determined by the Eq.~(\ref{ks1.1}). From this 
it is clear that the surfaces $r=const$ are confocal ellipsoids. While from Eq.~(\ref{ks1.2}), it follows that the 
$\theta=const$ surfaces are hyperboloids of one chart and confocal to ellipsoids.  

Sometimes the metric in Eq.~(\ref{ks18}) {could be}  written in the following form:
\begin{eqnarray}
 g_{ab} &=& \eta_{ab}+h\,k_{a}k_{b},
\end{eqnarray}
where $\eta_{ab}$ is the metric of flat space, $k_{a}$ is a null vector and $h$ is a scalar 
field as mentioned earlier. Since $k_{a}$ is defined as a null vector with respect to the background 
metric i.e., $k^a\equiv \eta^{ab}k_{b}=g^{ab}k_{b}$ and therefore $g^{ab}k_{a}k_{b}=0$, this indicates 
that $\eta^{ab}k_{a}k_{b}=0$. Kerr and Schild~\cite{ks65} proved that the vacuum Einstein equations 
should require that the null congruence to be geodesic with respect to the background metric $g_{ab}$ 
or $\eta_{ab}$. If the value of $h$ is chosen properly then $k_{a}$ should be 
normalized such that $k^b\nabla_{b}k^{a}=0$, where $\nabla_{b}$ is covariant derivative with 
respect to $g_{ab}$. Let us define $k^{a}=\frac{dx^{a}}{d\lambda}$ where $\lambda$ is an affine 
parameter then as Kerr and Schild deduced that 
\begin{eqnarray}
 R_{abcd}k^{b}k^{d} &=&- \left(\frac{d^2h}{d\lambda^2}\right) k_{a}k_{c}~.\label{ks21}  
\end{eqnarray}
It suggests that $k^{a}$ is a multiple Debever-Penrose~\cite{debever} vector and the Goldberg-Sachs~\cite{gs62}
theorem implies that the null geodesic congruence is defined by the $k^{a}$ shear-free. This  is valid for 
Kerr-MOG black hole also. The metric is of type II in the Petrov-Pirani classification as well. 

It is noted that in Eq.~(\ref{ks18}),  $r^2$ is defined implicitly in terms of $x, y, z$ by 
\begin{eqnarray}
r^4-(x^2+y^2+z^2-a^2)r^2-a^2z^2 &=& 0 ~.\label{ks22}
\end{eqnarray}
It is evident from Eq.~(\ref{ks18}), the metric is clearly analytic everywhere except at 
\begin{eqnarray}
x^2+y^2+z^2 &=& a^2\,\, \mbox{and}\,\, z=0 ~.\label{ks23}
\end{eqnarray}
otherwise it has a ring singularity. It is interesting to note that Eq.~(\ref{ks22}) is \emph{independent of 
the MOG parameter ($\alpha$)}. To see the curvature singularity in Kerr-MOG black hole, we have to compute the 
Kretschmann scalar as
\begin{eqnarray}
{\cal K} &=& {\cal R}_{abcd}{\cal R}^{abcd}= 
\frac{8 G_{N}^2{\cal M}^2}{(r^2+a^2\cos^2\theta)^6}f(r,a,\theta,\alpha), 
\end{eqnarray}
where the function $f$ is a function of $r,a,\theta$ and $\alpha$ only 
and $f(r,a,\theta,\alpha)\neq 0$ .
It is now clear that at $r=0$ and  $\theta=\frac{\pi}{2}$, the Kretschmann scalar diverges meaning that the 
only curvature singularity is present there. 

To understand the genuine nature of curvature singularity, one must eliminate the $\theta, \phi$ coordinates 
at $r=0$.  {The} Cartesian coordinates are awkward to study the global feature of Kerr-MOG 
black hole. But it is suitable to study the singular nature of the spacetime. From Eq.~(\ref{ks1.5}), it is 
evident that the surfaces $r=const$ are confocal ellipsoids as we have said earlier, whose principal axes 
coincide with the coordinate axes. These ellipsoids are degenerate, for $r=0$, to the disc 
{ i.e. }
\begin{eqnarray}
 x^2+y^2 \le a^2, z=0. 
\end{eqnarray}
Then the point $r=0, \theta=\frac{\pi}{2}$ corresponds  to the ring 
\begin{eqnarray}
 x^2+y^2=a^2, z=0  
\end{eqnarray}

This is the singularity structure of the Kerr-MOG black hole. The singularity along the 
ring is called ring singularity, since the Kretschmann scalar diverges there. 
It is interesting that it is independent of the MOG parameter $\alpha$. Note also that 
the above metric is nonsingular inside the ring i.e., at $r=0$ and  $\theta\neq \frac{\pi}{2}$ or 
\begin{eqnarray}
r=0,\,\,\, x^2+y^2 <a^2.   
\end{eqnarray}
The maximal analytic extension of the Kerr-MOG black hole exhibits characteristics that are quite 
analogous to those of the Kerr black hole. There are three distinct regions within the Kerr-MOG black 
hole: { $Region~ I: r_{H}< r <\infty$}, $Region~ II: r_{C}< r< r_{H}$
and $Region~ III:0< r <r_{C}$. In the region $r>r_{H}$ and $r<r_{C}$, $\Upsilon_{r}$ 
becomes $\Upsilon_{r}>0$. While in the region $r_{C}< r< r_{H}$, $\Upsilon_{r}$ becomes $\Upsilon_{r}<0$. 
The parameters $r_{H}$ and $r_{C}$ are denoted as outer horizon or event horizon and Cauchy horizon 
or inner horizon, respectively, as previously mentioned.

The Eq.(\ref{ks22}) yields four solutions, of which two are real and two are complex. The two real solutions correspond 
to the coordinate branches where $r>0$ and $r< 0 $. This indicates that for every set of values of the Cartesian 
coordinates $ x, y, z, t $, there are two distinct points of $ r $: one in the 
positive branch $ r > 0 $ and the other in the negative branch $r<0$. Consequently, we obtain two asymptotically 
flat spacetimes characterized by the metric Eq.(\ref{ks22}), one with $r > 0 $ and the other with $ r < 0 $. It 
is important to note that the asymptotic flat spacetime corresponding to $ r < 0$ does not possess any horizon.

{Now we} can  examine a timelike geodesic along the z-axis, which corresponds to a straight line in
the $(x,y,z)$ coordinate system. In the $(r,\theta)$ coordinates, this geodesic is represented by the 
equations $\frac{dr}{d\tau}=\pm\sqrt{E^2-\lambda^2}$ and $\frac{d\theta}{d\tau} \sim \sin\theta = 0$
{[See Ref. \cite{c1} for the geodesics of symmetry axis]}. 
Here, $E$, $\lambda$, and \(\tau\) denote the energy, mass, and proper time of the particle, respectively. 
Given that \(r\) must remain non-negative, it is necessary to select opposite signs for \(\sqrt{E^2 - \lambda^2}\) 
and to assign different values to \(\theta\) (either 0 or \(\pi\) for this geodesic) on the opposite side of 
the disk. Consequently, \(r(\tau)\) exhibits a lack of smoothness, while \(\theta(\tau)\) experiences discontinuity 
across the disk.

The metric described in equation (\ref{ks6}) can be geodesically extended to encompass negative values of \( r \).
Specifically, if \( r \) is permitted to take on negative values, there is no necessity to change the sign in front 
of \( \sqrt{E^2 - \lambda^2} \) in the aforementioned equation across the disk; thus, \( r(\tau) \) remains smooth. 
Furthermore, as will be demonstrated shortly, there is no requirement for \( \theta(\tau) \) to exhibit any 
discontinuities. Consequently, the geodesics can be analytically continued from the region where \( r < 0 \) to 
the region where \( r > 0 \). This principle applies not only to geodesics along the \( z \)-axis but also to all 
geodesics that do not intersect the ring located at \( (\rho, z) = (a, 0) \). Therefore, the metric in 
equation (\ref{ks6}) naturally extends to the interval \( r \in (-\infty, +\infty) \).

\section{Wormhole in Cylindrical Coordinates}

The metric described in equation \eqref{ks7.1} could represent a flat spacetime in $(\rho,z)$ coordinates. 
However, a notable distinction arises as the coordinate transformation \eqref{ks7.2} is no longer bijective. 
The pairs $(r,\theta)$ and $(-r,\pi-\theta)$ correspond to the same $(\rho,z)$ values, which implies that 
as $r$ and $\theta$ take on all possible values, $\rho$ and $z$ will each cover their ranges twice. 
Consequently, it is necessary 
to utilize two distinct $(\rho,z)$ coordinate patches to fully encompass the spacetime, which we shall 
denote as $(\rho_{+},z_{+})$ and $(\rho_{-},z_{-})$. In both coordinate patches, the metric retains the 
form given in 
\eqref{ks7.1}, but it could be expressed as follows:
\begin{eqnarray}
r+ia\cos\theta=+\sqrt{\rho_{+}^2+(z_{+}+ia)^2},\,\,r+ia\cos\theta=-\sqrt{\rho_{-}^2+(z_{-}+ia)^2}.
\end{eqnarray}
Thus, one coordinate patch corresponds to the Riemann sheet where $r>0$, while the other pertains to the 
sheet where $r<0$. Together, these coordinate patches encompass the entirety of the Riemann surface.

Each Riemann sheet features a branch cut along the segment \eqref{ks6.1}, with the two sheets being connected along these 
cuts by matching the upper side of one cut to the lower side of the other, and vice versa (See Fig.1 of  Ref.\cite{GW17}). 
Consequently, the $\theta$-coordinate transitions smoothly when moving from one coordinate patch to another, while the 
$r$-coordinate crosses zero and changes its sign.

Accordingly, the structure of spacetime can be understood as comprising two copies of \( R^4 \) interconnected 
via a disk, thereby forming a wormhole with two asymptotic regions. Locally, the geometry is flat and the curvature is zero; 
however, it is not globally flat due to the presence of a physical singularity at the ring located at 
\( (\rho_\pm,z_\pm)=(a,0) \). This can be demonstrated by observing that a contour encircling the branch point 
\( (a,0) \) in the \( (\rho_+,z_+) \) coordinate system does not return to its starting point after a single \( 2\pi \) 
revolution. Instead, it transitions to the \( (\rho_-,z_-) \) coordinate system and only after an additional \( 2\pi \) 
revolution does it return to the original chart, thus completing the loop. Consequently, the total angular increment 
amounts to \( 4\pi \), resulting in a negative angle deficit of \( 2\pi - 4\pi = -2\pi \), which indicates the presence 
of a conical singularity in the curvature at the ring.

One can reach the same conclusion by utilizing the $(r,\theta)$ coordinates. By defining \( r=a\,x_{1} \) and 
\(\theta= \cos^{-1} (x_{2}) \), the metric \eqref{ks6} becomes
\begin{eqnarray}
ds^2 &=& -dt^2 +a^2\Big(x_1^2 + x_2^2\Big)\Big[\frac{dx_1^2}{1+x_{1}^2} + \frac{dx_2^2}{1-x_{2}^2}\Big]
+a^2(1+x_{1}^2)(1-x_{2}^2)d\phi^2.
\end{eqnarray}
Analogously, the Jacobian of the coordinate transformation is 
\begin{eqnarray}
J &=& \Big|\frac{\partial(r,\theta)}{\partial (x_{1},x_{2})}\Big|\\
  &=&-{\frac{a}{\sqrt{1-x_{2}^2}}}, \,\, x_{2} \neq 1.
\end{eqnarray}
Interestingly the Jacobian is independent of coordinate $x_{1}$.  It becomes negative when $a>0$ and positive when $a<0$. 
{However,} for small values of \( x_1 \) and \( x_2 \), the above metric becomes
\begin{eqnarray}
ds^2 &=& -dt^2 +a^2 \Big(x_1^2 + x_2^2\Big)\Big[dx_1^2 + dx_2^2\Big] + a^2 d\phi^2
\end{eqnarray}
{Now if we apply  the coordinate} transformation as
\begin{eqnarray}
x_{1} &=& \sqrt{\frac{2x}{a}}\cos\Big(\frac{\theta}{2}\Big),\\
x_{2} &=& \sqrt{\frac{2x}{a}}\sin\Big(\frac{\theta}{2}\Big).
\end{eqnarray}
{Then the} metric contains a conical singularity at $x=0$ as
\begin{eqnarray}
ds^2 &=& -dt^2 + dx^2 + x^2 d\theta^2 + a^2 d\phi^2, 
\end{eqnarray}
The Jacobian of the above coordinate transformation is 
\begin{eqnarray}
J &=& \Big|\frac{\partial (x_{1},x_{2})}{\partial(x,\theta)}\Big|\\
  &=&\frac{1}{2a},\,\,\, a\neq 0.
\end{eqnarray}
The Jacobian is thus independent of coordinates: $x$ and $\theta$. Given that \(\theta \in [0, 4\pi) \), 
the metric exhibits a conical singularity at \(x = 0\), extending along the azimuthal direction of \(\phi \) 
[See the work~\cite{soko77} for ``bent conical singularities'']. The origin of the curvature singularity arises 
from an infinitely slender cosmic string characterized by negative tension~\cite{GW16,Vol17}. This cosmic 
string, which is structured as a ring, creates a void in spacetime, functioning as a portal that links the 
positive universe ($r>0$) with the negative universe ($r<0$).

\section{Zero Mass Limit of Kerr-MOG Black Hole is a Wormhole}
In this section, we {would like to} analyze the zero mass limit of Kerr-MOG black hole, which 
{could be  manifested} as either flat Minkowski spacetime or a wormhole characterized by locally 
flat geometry. To facilitate this examination, we have utilized the metric outlined in Eq.~\eqref{ks5}. Carter 
has demonstrated that the radial coordinate in this context can take on both positive and negative values, 
specifically $r\in(-\infty,\infty)$, leading to two asymptotic regions corresponding to $r\to\pm\infty$. 
The mass of the black hole exhibits opposite signs when observed from these two distinct regions. 
The curvature invariant is
\begin{eqnarray}\label{curv}
{\cal R}_{abcd}{\cal R}^{abcd}=
C_{abcd}C^{abcd}
=\frac{8 G_{N}^2{\cal M}^2}{(r^2+a^2\cos^2\theta)^6}f(r,a,\theta,\alpha), 
\end{eqnarray}
where 
$$
f(r,a,\theta,\alpha) = 6\Big(r^6-15a^2r^4\cos^2\theta+15a^4r^2\cos^4\theta-a^6\cos^6\theta\Big)
$$
\begin{eqnarray} 
-12G_{N}{\cal M}r\Big(\frac{\alpha}{1+\alpha}\Big)(r^4-10a^2r^2\cos^2\theta+5a^4\cos^4\theta)+\nonumber\\
\Big(\frac{\alpha}{1+\alpha}\Big)^2G_{N}^2{\cal M}^2(7r^4-34a^2r^2\cos^2\theta+7a^4cos^4
\theta\Big).
\end{eqnarray}
It diverges at $r=0$ and $\cos\theta=0$, which corresponds to a ring in the equatorial  plane. The singularity is 
covered by the horizon if 
$\frac{G_{N}^2{\cal M}^2}{1+\alpha} > a^2$ and is naked if $\frac{G_{N}^2{\cal M}^2}{1+\alpha}<a^2$.  
The geodesics that are not situated within the equatorial plane bypass the ring singularity and transition from the 
region where $r>\Big(\frac{\alpha}{1+\alpha}\Big)\frac{G_{N}{\cal M}}{2}$
to the region where $r<\Big(\frac{\alpha}{1+\alpha}\Big)\frac{G_{N}{\cal M}}{2}$. 
For instance, we {could} consider the timelike 
geodesics that align with the symmetry axis, which {could} be characterized by the following 
description.
\begin{eqnarray} \label{geod}
\frac{1}{\lambda^2}\left(\frac{dr}{d \tau}\right)^2+V_{eff}(r)=E~~~~~\mbox{with}~~~~~~V_{eff}(r)=-\frac{1}{r^2+a^2}
\Big[2G_{N}{\cal M} r-\Big(\frac{\alpha}{1+\alpha}\Big)G_{N}^2{\cal M}^2 \Big],
\end{eqnarray} 
where $E={\cal E}^2/\lambda^2-1$. The potential $V_{eff}(r)$ is attractive 
for $r>\Big(\frac{\alpha}{1+\alpha}\Big)\frac{G_{N}{\cal M}}{2}$ and repulsive for 
$r<\Big(\frac{\alpha}{1+\alpha}\Big)\frac{G_{N}{\cal M}}{2}$ and is perfectly regular 
at $r=0$, i.e.  
\begin{eqnarray} 
V_{eff}(r)|_{r=0}=\Big(\frac{\alpha}{1+\alpha}\Big)\Big(\frac{G_{N}^2{\cal M}^2}{a^2}\Big)  
\end{eqnarray}
If $E$ is larger than the maximum value of the potential,  
$$
V_{\rm max}=
\frac{G_{N}{\cal M}}{\Big[\sqrt{a^2+\Big(\frac{\alpha}{1+\alpha}\Big)^2\frac{G_{N}^2{\cal M}^2}{4}}
-\Big(\frac{\alpha}{1+\alpha}\Big)\frac{G_{N}{\cal M}}{2} \Big]},
$$,
then $r(\tau)$ interpolates over the whole range,  $r\in(-\infty,+\infty)$. 

Let us select non-zero values for $a$ and $\alpha$, and consider the limit as ${\cal M}$ approaches zero. In this 
situation, the potential \( V(r) \) uniformly approaches zero, allowing the particle to move freely within the 
interval \( r \in (-\infty, +\infty) \). As a result, the Kerr-MOG metric described in equation \eqref{ks5} simplifies 
in this limit to equation \eqref{ks6}, which characterizes a locally flat wormhole rather than the flat Minkowski 
spacetime that is typically assumed in existing literature.

It is important to highlight a specific point here. As previously discussed, the geometry represented by \eqref{ks6}
is locally flat; however, this characteristic alone does not suffice to ascertain its global structure. It is also 
necessary to define the range of the radial coordinate $r$. If one is examining the local geometry within an open 
set, such as for $r\in(0,\infty)$, it is accurate to assert that the limit of the Kerr-MOG metric as ${\cal M}\to 0$ 
is flat. Nevertheless, one cannot arbitrarily elect the spacetime topology when taking this limit. The original 
Kerr-MOG spacetime features two asymptotic regions, and the geodesics connecting them traverse the entire
interval, $r\in(-\infty,\infty)$. 

These characteristics must also be preserved in the limit as ${\cal M}\to 0$, indicating that the topology remains 
non-trivial in this scenario and corresponds to the aforementioned wormhole. It is worth noting that 
this spacetime still exhibits a curvature singularity.

The presence of a curvature singularity as ${\cal M} \to 0$ was notably highlighted by Carter in his work~\cite{c2}. 
He argued that in the specific scenario where ${\cal M}$ approaches zero, ``there must still be 
a curvature singularity at $r=\cos\theta=0$, even though the metric remains flat in all other regions''. More 
precisely, the curvature is composed of both the Weyl and Ricci components. As indicated in \eqref{curv}, the Weyl 
component diminishes as ${\cal M} \to 0$, while the Ricci tensor remains zero outside the singularity due to the 
vacuum nature of the metric. Nevertheless, it is permissible for the Ricci tensor to possess a non-zero value at 
the singularity, even in the limit of ${\cal M} \to 0$. As previously discussed, the 
metric \eqref{ks6} exhibits a conical singularity at the ring, which implies that the Ricci tensor has a 
delta-function structure concentrated at that ring. This phenomenon represents the continuation of the black hole 
ring singularity as ${\cal M} \to 0$.

An alternative method to demonstrate the same concept is to revert to a finite value of ${\cal M}$ and represent 
the Kerr-MOG metric using Kerr-Schild coordinates as derived in \eqref{ks18}.It is observed that the variables
$(\rho,z)$ in these equations correspond to $(r,\theta)$ in the same manner as indicated in \eqref{ks5}, thus 
the analysis presented in Sec. 3 is directly applicable. Consequently, given that $r\in(-\infty,+\infty)$, it is 
necessary to utilize two Kerr-Schild coordinate patches, namely $(\rho_{+},z_{+})$ and $(\rho_{-},z_{-})$, to 
adequately cover the manifold. Each coordinate patch features a branch cut at $\rho\in[0,a]$ and $z=0$. To 
achieve analytic continuation from one coordinate patch to the other, the upper side of one cut is matched 
with the lower side of the other, and vice versa. 

The Kerr-MOG black hole with \emph{superspinar} cases contains closed timelike curves (CTC). It can be readily examined 
with the $g_{\phi\phi}$ component of the metric described in \eqref{ks1}. 
\begin{eqnarray} 
g_{\phi\phi}= \Big[r^2+a^2+\frac{\Pi_{\alpha}}{\varrho^2} a^2\sin^2\theta\Big]\sin^2\theta.  
\end{eqnarray}
In the region where $r<\Big(\frac{\alpha}{1+\alpha}\Big)\frac{G_{N}{\cal M}}{2}$, which is near the ring, this 
component becomes negative, resulting in the closed orbits of the vector $\partial_{\phi}$ being timelike. These 
closed-timelike-curves {CTCs} must be modified to traverse any point within the spacetime continuum. However, in 
the scenario where ${\cal M}\rightarrow 0$, the $g_{\phi\phi}$ component remains positive, thus eliminating the issue.

\section{Is the Zero Mass Limit of the Kerr-Newman Black Hole a Wormhole?}

In this section, we will examine ``Does the zero mass limit of a Kerr-Newman black hole produce a wormhole?''. 
We will write the metric of a Kerr Newman black hole in the form:
\begin{eqnarray}
ds^2 &=& -dt^2+\varrho^2 \, \left(\frac{dr^2}{\Delta}+d\theta^2\right)+(r^2+a^2)\sin^2\theta d\phi^2
+\frac{(2mr-q^2)}{\varrho^2}\,\left(dt-a\sin^2\theta d\phi \right)^2 ~.\label{kn5},
\end{eqnarray}
where 
\begin{eqnarray}
\Delta &=& r^2-2mr+a^2+q^2.
\end{eqnarray}
As usual, the radial coordinate has both positive or negative, $r\in(-\infty,\infty)$, and there are two asymptotic 
regions corresponding to $r\rightarrow \pm\infty$ [See recent review~\cite{Adamo14,Hehl15,Teu15}].
Consequently, the black hole mass~($m$) and charge ($q$) must have two values in that regions. The curvature invariant
becomes the following: 
\begin{eqnarray}        
\label{curv1}
{\cal R}_{abcd}{\cal R}^{abcd}=
C_{abcd}C^{abcd}
=\frac{8}{(r^2+a^2\cos^2\theta)^6}g(r,a,\theta), 
\end{eqnarray}
where 
$$
g(r,a,\theta) = 6m^2\Big(r^6-15a^2r^4\cos^2\theta+15a^4r^2\cos^4\theta-a^6\cos^6\theta\Big)
$$
\begin{eqnarray} 
-12mq^2r\Big(r^4-10a^2r^2\cos^2\theta+5a^4\cos^4\theta\Big)+
q^4\Big(7r^4-34a^2r^2\cos^2\theta+7a^4cos^4\theta\Big).
\end{eqnarray}
Similarly, the curvature invariant  diverges at $r=0$ and $\theta=\frac{\pi}{2}$, which corresponds to a ring in
the equatorial plane. The singularity is covered by the horizon {when} $m^2>a^2+q^2$ and {the} 
superspinar {for} $m^2<a^2+q^2$.

Now we will see the geodesics which do not belongs to the equatorial plane miss the ring singularity and pass from the 
$r>q^2/2m$ region to the region $r<q^2/2m$. If we consider the timelike geodesics along the symmetry axis, then we will find 
\begin{eqnarray} \label{geod1}
\frac{1}{\lambda^2}\left(\frac{dr}{d \tau}\right)^2+V_{ef}(r)=E,
\end{eqnarray} 
where $E={\cal E}^2/\lambda^2-1$. Where the effective potential $V_{ef}$ as 
\begin{eqnarray}
V_{ef}(r) =-\frac{(2mr-q^2)}{r^2+a^2}.
\end{eqnarray}
Like Kerr-MOG black hole, the potential is attractive for $r>q^2/2m$ and repulsive for $r<q^2/2m$.  It is
perfectly regular at $r=0$, as determined by  
\begin{eqnarray} 
V_{ef}(r)|_{r=0}=\frac{q^2}{a^2}  
\end{eqnarray}
To determine the maxima or minima of the effective potential, we have to calculate  
$\frac{dV_{ef}}{dr}$:  
\begin{eqnarray} 
\frac{dV_{ef}(r)}{dr}=\frac{2m}{(r^2+a^2)^2}\left[r\left(r-\frac{q^2}{m}\right)-a^2\right]. 
\end{eqnarray}
To find the maximum value of the effective potential, we have to calculate the second derivative of the effective potential: 
\begin{eqnarray} 
\frac{d^2V_{ef}(r)}{dr^2} =\frac{2m}{(r^2+a^2)^3}\left[6a^2\left(r-\frac{q^2}{6m}\right)-2r^2\left(r-\frac{3q^2}{2m}\right)\right]. 
\end{eqnarray}
Let us find the maximum value of the effective potential  {occurs} at $r=\frac{q^2}{2m}-\sqrt{\frac{q^4}{4m^2}+a^2}$. If $E$ is larger 
than the maximum value of the potential, then maximum value of the potential is
$$
V_{\rm max}=\frac{m}{\Big[\sqrt{a^2+\frac{q^4}{4m^2}}-\frac{q^2}{2m}\Big]},
$$
and $r(\tau)$ interpolates over the whole range, $r\in(-\infty,+\infty)$. 

In a manner similar to Kerr black holes~\cite{GW17} or Kerr-MOG black holes, we consider the scenario 
where \( a \neq 0 \) and examine the limit as \( m \rightarrow 0 \). Under these conditions, the effective 
potential \( V_{ef}(r) \) consistently 
approaches a non-zero value. It becomes zero when the charge \( q \) approaches zero, thereby permitting the particle to 
traverse freely within the range \( r \in (-\infty, +\infty) \). Consequently, the Kerr-Newman metric presented in 
equation \eqref{kn5} simplifies in this limit to equation \eqref{ks6}, which characterizes a locally flat wormhole 
instead of the flat Minkowski spacetime. This represents another intriguing outcome of this study. 

Therefore, \emph{zero mass limit of Kerr-Newman black hole is not a wormhole}. It transforms into a wormhole only 
when an additional condition on the charge \( q \) is imposed, specifically when \( q = 0 \). Under this condition, the 
Kerr-Newman black hole can be classified as a wormhole. Hence, we conclude that \emph{zero mass and zero charge limit 
of Kerr-Newman black hole is a wormhole}. 

Another remarkable point to be noted is that  when $m\rightarrow 0$, the curvature invariant is not equal to zero. It 
will be zero when we adopt the additional condition $q=0$.
In fact, when $m=0$ and $q=0$ there must still be a curvature singularity at $r=0$ and 
$\theta=\frac{\pi}{2}$. However, the metric is then flat everywhere. It should be noted that the curvature invariant 
consists 
of two parts. One is the Weyl part and another is the Ricci part. The Weyl part vanishes as $m\rightarrow 0$ and 
$q\rightarrow 0$ as seen from \eqref{curv1}, as the Ricci tensor is zero outside the singularity since the metric 
is vacuum in nature. It is important to note that the Ricci tensor may indeed have a non-zero value at the 
singularity, even as the parameters $m$ and $q$ approach zero. As previously mentioned, the metric \eqref{ks6} 
displays a conical singularity at the ring, indicating that the Ricci tensor takes on a delta-function form that 
is localized at that ring. This occurrence signifies the persistence of the black hole ring singularity as $m\rightarrow 0$ 
and $q\rightarrow 0$.

An alternative approach to illustrate the same concept involves reverting to a finite value of $m$, $q$ and expressing 
the Kerr-Newman metric in Kerr-Schild coordinates~\cite{ks65}, which is related to the Boyer-Lindquist 
coordinates $t,r,\theta,\phi$ in \eqref{kn5} via 
\begin{eqnarray} 
 x=\rho \cos\phi,\,\, y=\rho\sin\phi,\,\, z=r\cos\theta,\,\, T=t+\int\frac{2mr-q^2}{\Delta}dr,
\end{eqnarray}
where
\begin{eqnarray} 
\rho=\sqrt{x^2+y^2},\,\, \varphi=\phi+\int\frac{a(2mr-q^2)}{\varrho\Delta}dr,
\end{eqnarray}
which produces the metric 
\begin{eqnarray} 
ds^2=-dT^2 +dx^2+dy^2+dz^2+\frac{r^2(2mr-q^2)}{r^4+a^2z^2}
\left[dt+\frac{r(xdx+ydy)}{r^2+a^2}-\frac{a(xdy-ydx)}{r^2+a^2}+\frac{zdz}{r}\right]^2 
~.\label{kskn}
\end{eqnarray}
Here, $r$ is determined implicitly in terms of $x$, $y$, $z$, by 
$$
r^4-(x^2+y^2+z^2-a^2)r^2-a^2z^2=0.
$$
It is evident from Eq.~(\ref{kskn}) that when $m=0$, it does not reduce to Eq.~(\ref{ks6}). Instead, it reduces
to Eq.~(\ref{ks6}) only when both mass parameter and charge parameter become zero {i.e.}
$m=0$ and $q=0$.

The variables $(\rho,z)$ in the equations we previously discussed correspond to $(r,\theta)$ in a manner similar to 
that indicated in \eqref{kn5}. Therefore, the analysis provided in Section 3 is directly relevant. Given that $r$ 
spans the interval $(-\infty,+\infty)$, it is essential to employ two Kerr-Schild coordinate 
patches, specifically $(\rho_{+},z_{+})$ and $(\rho_{-},z_{-})$, to comprehensively cover the manifold. Each of these 
coordinate patches contains a branch cut at $\rho\in[0,a]$ and $z=0$. To facilitate analytic continuation 
between the two coordinate patches, the upper side of one cut is aligned with the lower side of the other, and vice versa.

Our investigation has concentrated on the Kerr-Newman black hole, with a particular emphasis on superspinar scenarios 
that exhibit closed timelike curves. This phenomenon can be effectively examined through the $g_{\phi\phi}$ 
component of the metric, as described in \eqref{kn5}. 
\begin{eqnarray} 
g_{\phi\phi}= \Big[r^2+a^2+\frac{\left(2mr-q^2 \right)}{\varrho^2} a^2\sin^2\theta\Big]\sin^2\theta.  
\end{eqnarray}
In the region where $r<\frac{q^2}{2m}$, which is located near the ring, this component becomes negative. As a 
result, the closed orbits of the vector $\partial_{\phi}$ are classified as timelike. It is essential to adjust 
these CTCs to facilitate traversal through any point within 
the spacetime continuum. However, in the scenario where $m\rightarrow 0$ and $q\rightarrow 0$, the $g_{\phi\phi}$ 
component remains positive, thus addressing the concern.

\section{Conclusion}

It has been argued in the literature~\cite{GW17} that the zero mass limit of a spinning uncharged black hole does 
not correspond to flat Minkowski spacetime, but rather to a spacetime characterized by local flatness in its 
geometry. This limiting spacetime is referred to as the spinning uncharged spacetime. It features two asymptotic 
coordinate regions and possesses a topologically non-trivial structure. Additionally, it includes a curvature 
singularity. While the Weyl component of the curvature approaches zero as \( M \rightarrow 0 \), the distributional 
contribution of the Ricci tensor, which is supported by the ring, does not diminish in this limit. Consequently, this 
spacetime can be interpreted as a wormhole generated by a negative tension ring. This interpretation also applies to 
the Kerr-AdS spacetime.

Inspired by this work, we studied and found that the zero mass limit of Kerr-MOG black hole is a wormhole. 
{To do so}, we {first}
derived the Kerr-Schild form of Kerr-MOG black hole, which  is characterized by three parameters: the mass parameter, the 
spin parameter, and the MOG parameter. {Then we} derived the Kerr-Schild form within the framework of MOG theory, analysed through 
a specific coordinate transformation. Like spinning uncharged spacetime this black hole also contains two asymptotic regions and 
it has non-trivial topological structure. Moreover, it possesses a curvature singularity. 

{Moreover, } we showed that as the Weyl component of the curvature {tend towards}
zero when \( {\cal M} \rightarrow 0 \), the 
distributional effect of the Ricci tensor, which is sustained by the ring, remains unchanged in this limit. 
Therefore, this spacetime can be understood as a wormhole created by a ring with negative tension. Additionally, we 
have examined the maximal analytic extension of the black hole.
Finally, we extended our analysis to the case of the Kerr-Newman black hole in which we showed that the zero mass 
limit of Kerr-Newman black hole is not a wormhole. However, if we impose the additional criterion that both the mass parameter 
and the charge parameter are equal to zero, then the Kerr-Newman black hole transforms into a wormhole. 

\section*{Acknowledgments}

Research at the Perimeter Institute for Theoretical Physics is supported by the Government of Canada through industry 
Canada and by the Province of Ontario through the Ministry of Research and Innovation (MRI).

\end{document}